\documentclass[prl,twocolumn,superscriptaddress,notitlepage,longbibliography]{revtex4-2}
\usepackage{amsmath}
\usepackage{physics}
\usepackage{amssymb}
\usepackage[T1]{fontenc}
\usepackage{booktabs}
\usepackage{tabularx}

\usepackage[unicode=true,colorlinks=true,citecolor=blue,urlcolor=blue]{hyperref}

\usepackage{bm}
\usepackage{xcolor}
\usepackage{epsfig} 
\usepackage{multirow}
\usepackage{soul}

\renewcommand {\Im}{\mathop\mathrm{Im}\nolimits}
\renewcommand {\Re}{\mathop\mathrm{Re}\nolimits}

\renewcommand {\phi}{{\varphi}}
\newcommand {\rmi}{{\rm i}}

\usepackage[normalem]{ulem}
\begin{document}

\title{Persistent subradiant correlations in a random driven Dicke model}

 \author{Nikita Leppenen}
\email{nikita.leppenen@weizmann.ac.il}
\affiliation{Department of Chemical \& Biological Physics, Weizmann Institute of Science, Israel, Rehovot 7610001 }

\author{Alexander N. Poddubny}%
\email{poddubny@weizmann.ac.il}

\affiliation{Department of Physics of Complex Systems, Weizmann Institute of Science, Israel, Rehovot 7610001}

\begin{abstract}
We study theoretically the driven-dissipative dynamics of an array of two-level emitters, coupled to a single photonic mode, in the presence of disorder in the resonant frequencies.  We introduce the notion of subradiant correlations in the dynamics,  corresponding to the eigenstates of the Liouvillian with a low decay rate, that can also oscillate in time. While the usual collective subradiant states do not survive the emitter resonant frequency fluctuations, these subradiant correlations are immune to such a type of disorder. 
These long-living correlations exist in finite-size systems, when their lifetime is parametrically longer than in the so-called Dicke time crystal phase.
\end{abstract}

\maketitle

The interaction between a quantum system and its environment induces dissipation, which depends on the interaction strength and the system parameters. Applying an external drive to the system can counteract the dissipation, leading to the emergence of non-equilibrium states. The interplay between driving and dissipation near exceptional points of the operators, governing the system dynamics, gives rise to novel quantum many-body phenomena, such as dissipative phase transitions~\cite{Werner2005,Capriotti2005,Kessler2012,Lee2014,Barberena2019}.  The exceptional points indicating these phase transitions can be found from the spectral analysis of the Liouvillian superoperator $\mathcal L$ that describes the open systems dynamics~\cite{Albert_2016,Minganti2018}. Depending on the type of the phase transition, its signature could be vanishing of the real part, also referred to as a Liouvillian gap, of one or several eigenvalues $\lambda$ of the superoperator ${\cal L}$~\cite{Wilson2016,Vicentini2018,leppenen2024}. 

\begin{figure}[t!]
    \centering
    \includegraphics[width=1\linewidth]{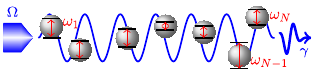}
    \caption{Sketch of the system. $N$ two-level atoms with resonant frequencies $\omega_n$ ($n = 1,\ldots,N$) are coupled to the same photon mode driven with the Rabi frequency $\Omega$. Atoms collectively emit to the mode with the decay rate $\gamma$.}
    \label{fig:schematics}
\end{figure}

In this Letter, we explore the Liouvillian points with vanishing real part in terms of the subradiant long living states and correlations for a driven-dissipative Dicke-type model, a two-level atom ensemble in the presence of disorder, see schematics in Fig.~\ref{fig:schematics}(a). In the absence of the disorder, this system features a superradiant phase transition~\cite{Carmichael_1980}. This effect  has become a focus of advanced theoretical~\cite{Efi_CRSS, Cabot_2023,Drummond1978} and experimental investigations~\cite{Ferioli2023,song2024}
and is now also referred to as dissipative or boundary time crystal phase~\cite{Iemini2018,Buca2019, Cabot_2023,solanki2024chaostimedissipativecontinuous}. 
However,  the strong fluctuations of the atom frequencies suppress the formation of collective superradiant and subradiant states, with enhanced or suppressed spontaneous decay rate in the absence of drive~\cite{Ferioli_2021}. 
 How these collective effects manifest in the driven Liouvillian dynamics and which type of disorder they survive is still an open question to the best of our knowledge.

Here, we demonstrate the presence of long-living {\it subradiant correlations} with small $|\Re \lambda|$, being a signature of collective dissipation, that are robust to the fluctuations of the atomic transition frequencies. We show that while collective interactions and subradiance are destroyed by the fluctuations for vanishing driving, they can be recovered for large driving strength.  Such recovery is related to the dynamical decoupling effect, well known for driven inhomogeneously broadened ensembles of quantum systems~\cite{Sagi_2010,Almog_2011,Wang_2012,Barry2020,covolo2025nonmarkoviandynamicscollectivelyencodedqubits,Retzker2025}. Contrary to the typical dynamical decoupling setup, here we focus on the rate of emission into the photonic bath, and not just on the dephasing in the ensemble. 
We rigorously find the symmetry conditions required for the slow collective decay and collective oscillations, where $\Im\lambda\ne 0$ and small $|\Re \lambda|$.  To this end, we propose a method to characterize the degeneracy of oscillations based on the group representation theory.  Both the oscillation frequency and lifetime are found to be very sensitive to the dipole-dipole interactions.

The predicted subradiant correlations qualitatively differ from the correlations in the so-called dissipative time crystals and quasicrystals~\cite{solanki2024chaostimedissipativecontinuous} that, while also offering a variety of new oscillation frequencies, are long-living only in the thermodynamic limit of $N\to \infty$ atoms.


{\it Theoretical model.}  We describe the driven-dissipative system of the $N$ atoms with different transition frequencies by a random Dicke model, illustrated in Fig.~\ref{fig:schematics}. The model dynamics is governed by the Liouvillian ${\cal L}$, defined as 
\begin{gather}
    {\cal L} \rho = \dot{\rho}  = -\rmi [{\cal H},\rho] + \frac{\gamma}{2}(2\hat{J}_-\rho \hat{J}_+ - \rho \hat{J}_+ \hat{J}_- - \hat{J}_+ \hat{J}_- \rho), \notag  \\ \label{eq:me} {\cal H} = 2\Omega \hat{J}_x+\delta H,\quad  \delta H= \sum_{n = 1}^N \omega_n \sigma_n^z.
\end{gather}
Here, we introduce the collective spin operators $\hat{J}_{x},\hat{J}_-$, constructed from the annihilation operators of individual two-level atoms, $\sigma_n$ ($n = 1,\ldots,N)$. These operators describe the collective decay jump $\hat{J}_- = \sum_{n}\sigma_n$ ($\hat{J}_+ = \hat{J}_-^\dagger$) and the collective coherent drive, $\hat{J}_x = 1/2\sum_n \sigma_n^x$. We assume identical couplings of all the atoms to the photonic mode. This allows us to consider only one collective jump operator $\hat{J}_-$ and identical driving strength $\Omega$ for all of the atoms.

\begin{figure}[t!]
\centering\includegraphics[width=\linewidth]{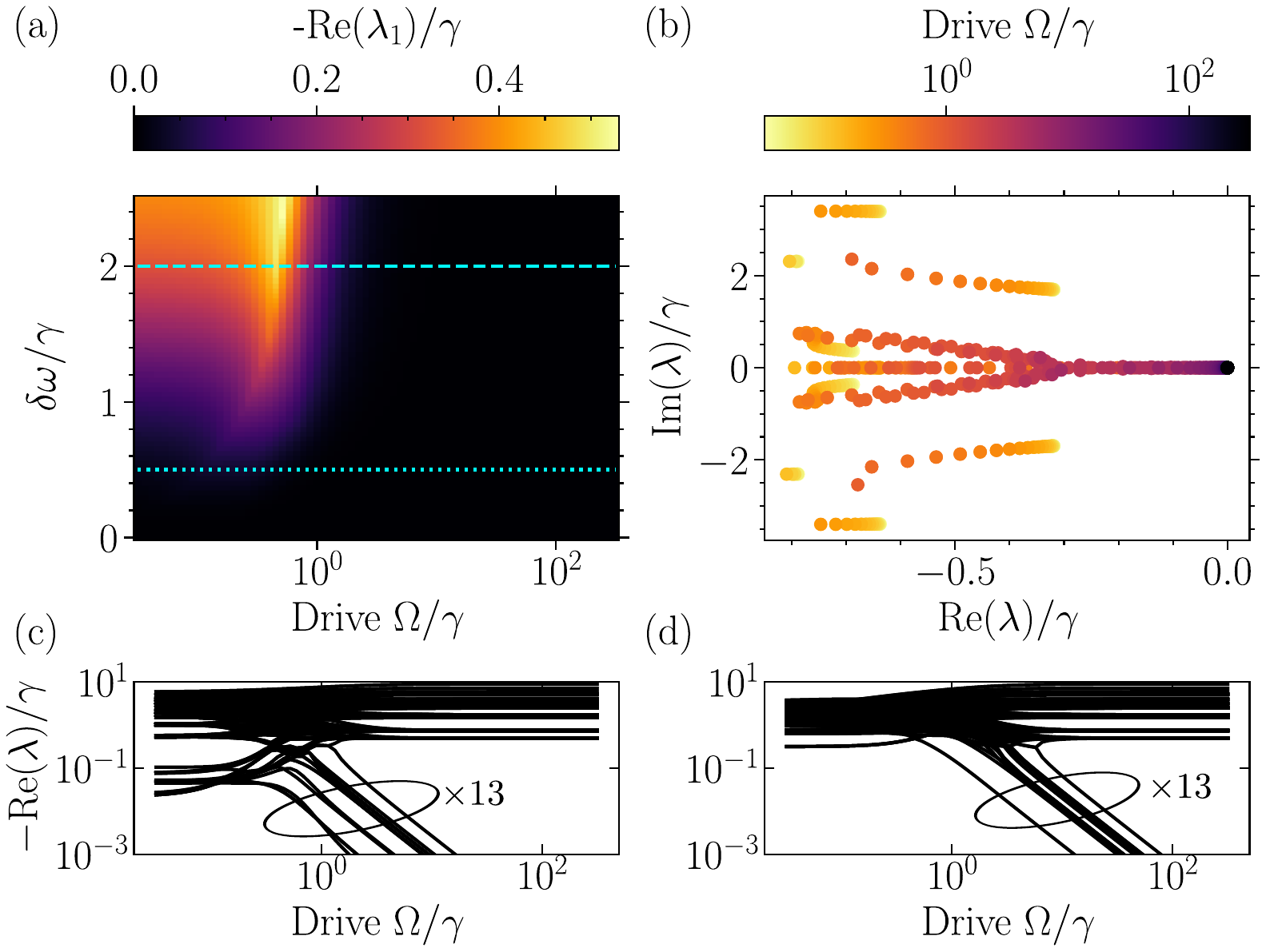}
\caption{(a) Phase diagram (a lifetime map shown in color depending on drive and disorder) for the smallest nonzero value of Liouvillian spectrum $\lambda_1$. (b) The scatter plot of the first 14 eigenvalues in the complex plane for the different drive values for $\delta \omega/\gamma = 2$.  (c,d) Dependence of the nonzero
Re $\lambda$  (without the steady state $\lambda = 0$)  on the driving strength for two different values of $\delta \omega/\gamma=\{2,0.5\}$, indicated on the panel (a) by thin horizontal dashed (d) [dotted (c)] lines.
Calculation has been performed for $N=4$.}
 \label{fig:map}
\end{figure}

The symmetric non-Hermitian matrix ${\cal L}$ acts in the space of the $N$ qubits density matrices and has the dimension $4^N \times 4^N$. 
It is the inhomogeneous broadening term $\sum_{n = 1}^N \omega_n \sigma_n^z$ in the Hamiltonian that breaks the permutational symmetry of the problem and prevents it from being described solely in terms of collective operators. As such, it is not possible to restrict the Hilbert space just to the $(N+1)$-dimensional Dicke manifold --- the set of individual transition frequencies $\omega_n$ appeared due to the assumed inhomogeneity of the ensemble.

We now introduce  the Liouvillian spectrum: 
\begin{equation}\label{eq:L_eig}
    {\cal L} \rho_i = \lambda_i \rho_i, \qquad \rho(t) = \sum_i c_i \rho_i e^{\lambda_i t}\:.
\end{equation}
The equation Eq.~\eqref{eq:me} has just one steady state solution $\rho_s$ with $\lambda\equiv \lambda_s\equiv  0$.  However, the non-trivial effects could occur in the time dynamics towards the steady state. As we assume the steady-state is reached at $t \to \infty$, that automatically requires $\text{Re}(\lambda_i)\leq 0$. One can then interpret $-1/\Re\lambda_i$ for any nonzero $\lambda_i$ as the lifetime of the eigenstate $\rho_i$, which, due to the zero trace, has the meaning of the correlation function~\cite{poddubny_2022_driven}.

The non-zero eigenvalues of ${\cal L}$ for $N = 4$ as a function of the normalized driven strength $\Omega/\gamma$ and the inhomogeneous broadening $\delta\omega$ are shown in Fig.~\ref{fig:map}.
Without the loss of generality, we have chosen a simplified model of the inhomogeneous broadening, where the eigenfrequencies $\omega_n$ are equidistantly spaced from $-\delta\omega$ to $\delta\omega$. The panel Fig.~\ref{fig:map}(a) presents by color the smallest nonzero value of Liouvillian spectrum $\lambda_1$. This can be interpreted as the decay rate of the longest-living nonstationary correlation in the system. The calculation demonstrates that when the driving is weak, $\Omega \lesssim \gamma$, the correlations decay faster for stronger inhomogeneous broadening. There are no correlations with lifetime longer than $1/\gamma$ for $\Omega \lesssim \gamma$ and $\delta\omega\gtrsim \gamma$ --- see the bright spot in the top-left corner of the map Fig.~\ref{fig:map}(a). The situation qualitatively changes for larger driving strengths. In this case, long-living correlations are present regardless of the broadening.

In order to illustrate this in more detail, we show in  Fig.~\ref{fig:map}(c,d) the dependences of all the decay rates $\Re \lambda(\Omega)$ for strong and weak inhomogeneous broadening, respectively. When the broadening is almost absent (c), the long-living correlations with $\Re \lambda\ll \gamma$ exist for any driving strength. Strong broadening (d) destroys the long-living correlations for weak driving, which are, however, recovered for strong driving. 
{The trend of the imaginary parts of the spectrum is seen in Fig.~\ref{fig:map}(b). As we increase the drive, both the imaginary and the real parts of the eigenvalues are suppressed.}

{\it Long living correlations.} Here, we provide a simple picture allowing us to explain the persistence of the long-living correlations in Fig.~\ref{fig:map} for a strong disorder as well as to calculate their number.

\begin{figure}
    \centering

    \includegraphics[width = \columnwidth]{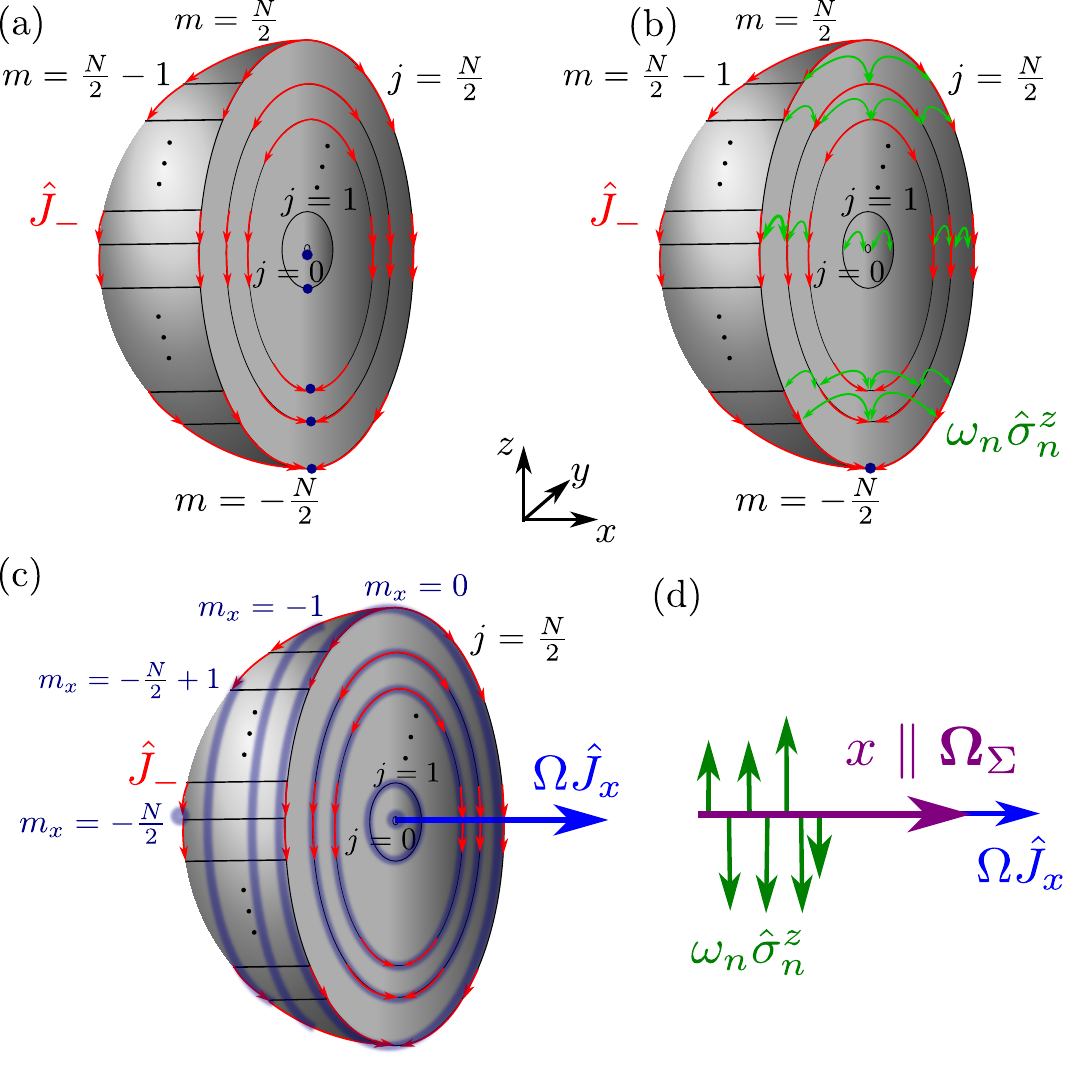}
    \caption{Representation of the dark states on the collective Bloch sphere. Blue orbits in the cross section correspond to the different $j$ manifolds, and black lines of latitude on the surface indicate states with different $m$-values.  The action of the collective jump $\hat{J}_-$ causes the transition to the lower $m$ value in each $j$ manifold and is shown by the red arrows. (a) Pure dark states in the non-driven Dicke model (dark blue dots) are located at the north pole of each $j$ manifold. (b) Inhomogeneous broadening (green double arrows) causes the transition between different $j$ manifolds and destroys the dark states except for the ground state. (c) An external drive term $\Omega \hat{J}_x$ is shown as an action of the effective magnetic field along the $x$ axis (blue arrow).
    Mixed dark states in the driven Dicke model (dashed blue orbits) align with the mixture of the $\hat{J}_x$ eigenstates within each $j$ manifold.  (d) The total field $\bm \Omega_\Sigma$ of the strong drive and effective magnetic fields due to the inhomogeneous broadening is oriented along the drive direction.}
    \label{fig:bloch}
    \end{figure}

 We use the Bloch representation of the density matrix, see Fig.~\ref{fig:bloch}. For each value of the collective spin $j$, we plot a corresponding Bloch sphere, and the overall state is described by the distribution of points on the nested spheres.
The jump operators $\hat{J_-}$, shown by the red arrows in Fig.~\ref{fig:bloch}(a), push the states to the south poles on corresponding spheres, but do not mix the spheres. Thus, the south poles are invariant to the jumps and correspond to dark states.  There exist $C_{N}^{\lfloor N/2\rfloor}$ such dark states, including the ground state~\cite{poshakinskiy_dimerization_2021}, that form  the kernel of the jump operator $\hat{J}_-$~\cite{Damanet2016,shammah_superradiance_2017} (see also Appendix~A). Those states are the states with $-\Re(\lambda)<\gamma$ that are seen at the small drive values in Fig.~\ref{fig:map}(c).

However,  the inhomogeneous broadening term $\delta H=\sum_n \omega_n \hat{\sigma}_n$ breaks the rotational symmetry. It can be interpreted as effective random magnetic fields along $z$ [green arrows in Fig.~\ref{fig:bloch}(d)], that trigger the transitions between states with $j\ne j'$ shown in Fig.~\ref{fig:bloch}(b).
 This would destabilize the points at the south poles and would have broken the subradiance if not for the external drive.  As can be seen from Eq.~\eqref{eq:me}, the external drive leads to the magnetic field along $x$-axis,  see blue arrows in Fig.~\ref{fig:bloch}(c,d). When drive overcomes the disorder ($\Omega \gg \delta \omega$), the total magnetic field tends to the $x$ direction, and the transitions between different $j$-th are suppressed.  This is the essence of the dynamical decoupling effect~\cite{Sagi_2010,Almog_2011,Wang_2012,Barry2020}.
 The long living states are now equally distributed (mixed states) at the orbits spinning about the $x$ axis and conserving $j$ [blue circles in Fig.~\ref{fig:bloch}(c)].
 Being rotationally invariant, these orbits are unaffected by the jumps and hence correspond to subradiant correlations.

More formally,  the orbits correspond to the uniform mixture of the angular momentum states quantized along $x$ axis ($\hat{J}_x \ket{j,m_x,\nu} = m_x \ket{j,m_x,\nu}$) at each $j$ manifold 
\begin{equation}\label{eq:rho_drive}
\rho(j)_d^{\nu,\nu'} = \frac{1}{2j+1}\sum_{m_x=-j}^j  \ket{j ,m_x ,\nu} \bra{j, m_x, \nu'}.
\end{equation}
The sum over $m_x$ leads to invariance of Eq.~\eqref{eq:rho_drive} to the jumps,
and the conservation of $j$ ensures stability to the disorder in the limit of $\Omega\gg\delta\omega\gg \gamma$.
The indices in Eq.~\eqref{eq:rho_drive} $\nu, \nu' = 1,..,d_j$ numerate different irreducible representations at each $j$. Due to the dynamical decoupling from disorder, the number of subradiant correlations for a strong drive is the same as in the absence of the  disorder~\cite{poddubny_2022_driven}. Thus, in the strongly driven case, we have 
\begin{equation}\label{eq:djdrive}
    \sum_j^{N/2}d_j^2 = \frac{4^N \Gamma\qty(N+\frac{1}{2})}{\sqrt{\pi}\Gamma(N+2)}
\end{equation}long-living correlations~\footnote{Dmitry Smirnov, private communication}. For $N = 4$, this formula gives 14 states, including the steady state, that is seen in the limit $\Omega \to \infty$ in Fig.~\ref{fig:map}(c-d).

\begin{figure}[ht!]
\centering
\includegraphics[width = \columnwidth]{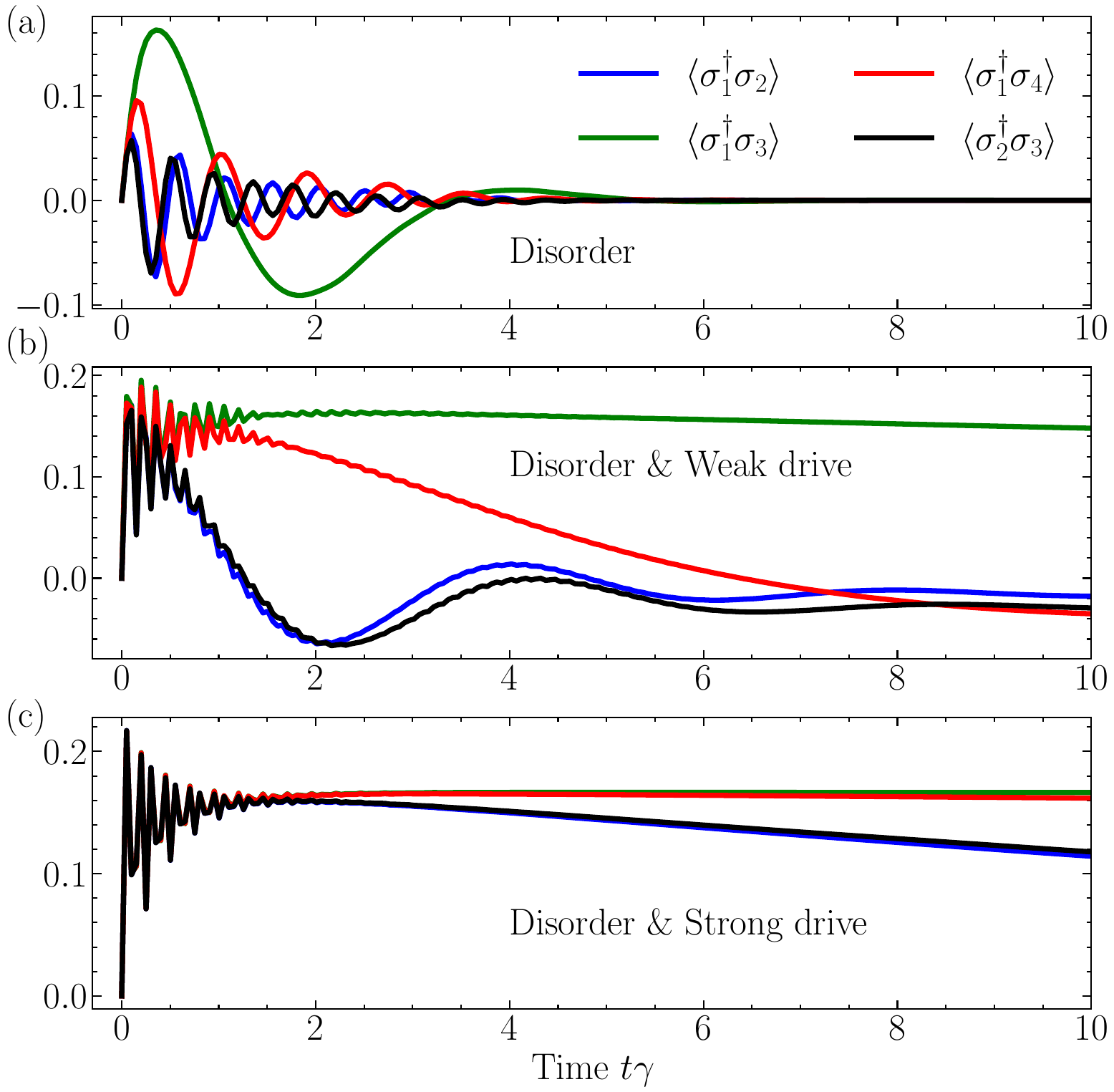}
\caption{Time dynamics of the correlation functions $\langle \hat{\sigma}_{i}^\dagger \hat{\sigma}_j \rangle$ for $N = 4$ in the case of different drive to dissipation relation (a) $\Omega = 0$, (b) $\Omega = 20 \gamma$, (c) $\Omega = 200 \gamma$. In the simulations, we used random disorder generated from the normal distribution with the scale $10\gamma$ (see Appendix~B for exact values of parameters).  } \label{fig:time_dyn}
\end{figure}

From Eq.~\eqref{eq:L_eig} it follows that long-lived correlations are observed in the temporal dynamics shown in Fig.~\ref{fig:time_dyn}. Indeed,  we clearly observe the appearance of the long-living states when the driving strength increases, compare panels (a) to (c). In Appendix~B, we show that these states correspond to the eigenvectors of the Liouvillian with a small real part.
While the real parts of the Liouvillian spectrum define their lifetimes, the existence of the non-zero imaginary parts indicates the oscillatory behaviour. However, as shown in Fig.~\ref{fig:map}(b), at the strong drive values, the imaginary part is reduced. This is consistent with the time dynamics: as we increase the driving strength, we see that the correlations' lifetime increases but the oscillations disappear [Fig.~\ref{fig:time_dyn}(b-c)]. We propose the model of the non-vanishing imaginary parts in the next section.

{\it Oscillating correlations.} We have shown that a large drive decouples the system from the disorder. Generally, this is not the case for an arbitrary Hamiltonian perturbation $\delta H$. For example, one could consider including in the model nearest-neighbor dipole-dipole interaction
\begin{equation}\label{eq:dd_nn}
\delta {\cal H}_{\rm NN} = \Delta \sum_{\rm n \in NN} ( \hat{\sigma}_n^\dagger \hat{\sigma}^{\vphantom{\dag}}_{n+1}+\text{H.c.}).
\end{equation}
Together with the Hamiltonian and dissipator in Eq.~\eqref{eq:me} this system reminds waveguide QED system~\cite{pichler2015quantum,poddubny2023RevModPhys} where the atoms located at the short distance $d$ from each other so that we take into account only first-order contributions in $d/\lambda$. 
The near-field inductive couplings also exist in realistic superconducting qubit setups~\cite{zanner2021coherent}.   

\begin{figure}[tb]
    \centering
\includegraphics[width=1\columnwidth]{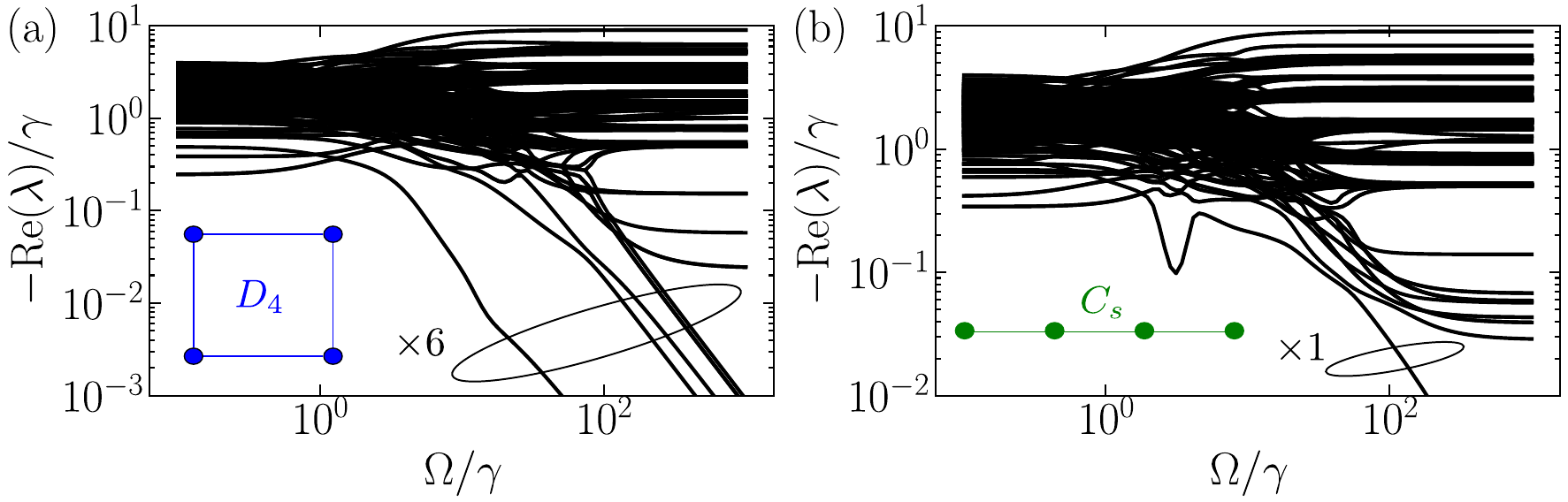}
\includegraphics[width=1
\columnwidth]{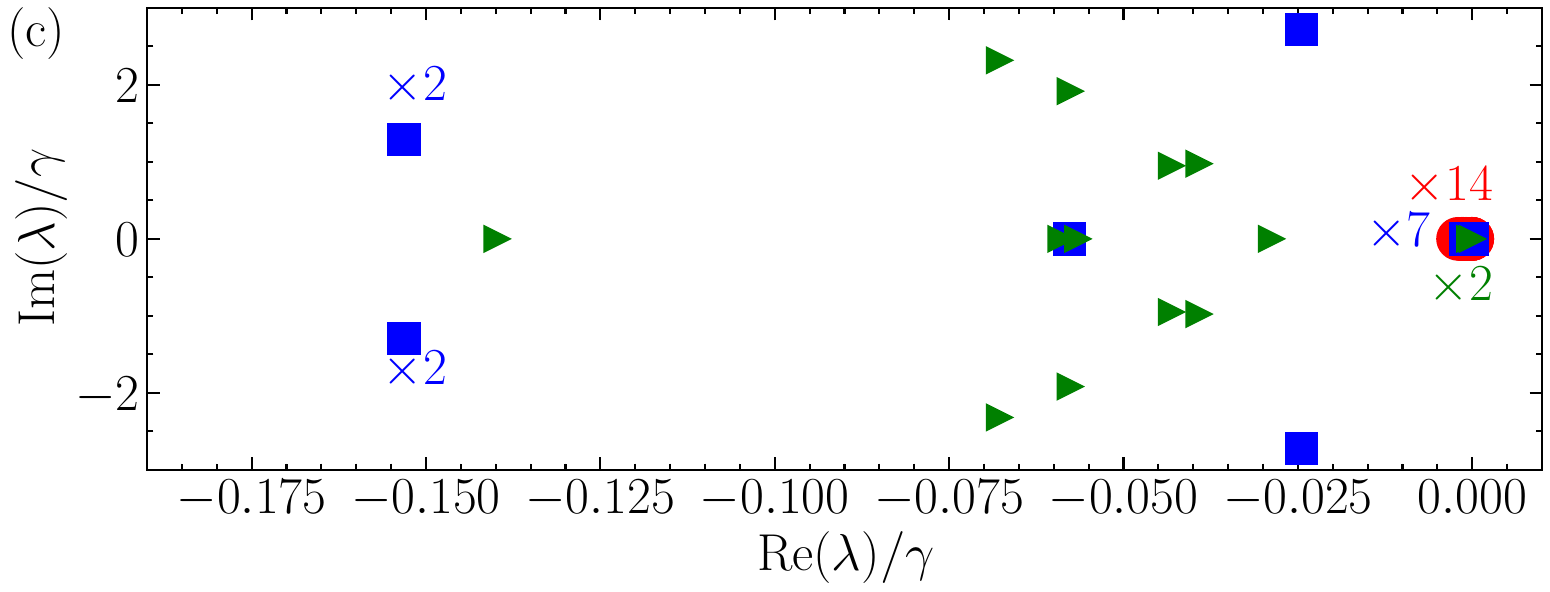}
    \caption{Dependence of the nonzero $\text{Re}(\lambda)$ (without the steady state $\lambda = 0$)  on the driving strength in the model Eq.~\eqref{eq:me} with the nearest-neighbour dipole-dipole interaction Eq.~\eqref{eq:dd_nn} for  two different geometries shown in the inset: (a) periodic boundary conditions corresponding to the $D_N$ symmetry group, (b) open boundary conditions corresponding to the $C_s$ symmetry group.  (c) The scatter plot of the first 14 (including the steady state $\lambda = 0$) Liouvillian eigenvalues at $\Omega = 200\gamma$ for the systems without dipole-dipole interaction that corresponds to the $S_N$ symmetry group (red dots); $D_N$ symmetry group (blue squares); $C_s$ symmetry group (green triangles). Calculation has been performed for $N = 4$, $\Delta = \gamma$ and disorder similar to Fig.~\ref{fig:time_dyn} (see Appendix~B for exact values).}
    \label{fig:dipole-dipole}
\end{figure}

We show in Fig.~\ref{fig:dipole-dipole}, that such Hamiltonian correction modifies the number and the behavior of the long-living correlations compared to the only drive and disorder case. Interestingly, it also depends on the boundary conditions in Eq.~\eqref{eq:dd_nn}. In order to understand this effect, we employ the group representation theory approach~\cite{bir1974symmetry,GelessusCharacterTables2025} explained in Appendix~A and summarized below.   

As shown in Fig.~\ref{fig:map}(c-d) for $N = 4 $ and $\Delta=0$, the spectrum in the limit of $\Omega\to\infty$ has 14 degenerate subradiant eigenvalues with $\Im\lambda=0$, within the ground state given by Eq.~\eqref{eq:rho_drive}. In contrast, with the dipole-dipole term with periodic boundary conditions, corresponding to the $D_N$ point symmetry, and with the open boundary, corresponding to the $C_s$ point symmetry, we have $7$ and $2$ dark states, respectively, in the limit of $\Omega\to\infty$. Generally, we found, that for any group and number of atoms, the number of the correlations with infinity long live time is given by $\sum_i \abs{\mathop\mathrm{dim}D_i}^2$, where $\mathop\mathrm{dim}D_i$ is the dimension of the representation $D_i$ that describes the transformation of part of the angular momentum states. For the symmetric group $S_N$, which corresponds to our system of $N$ atoms with $\Delta = 0$, this formula converges to Eq.~\eqref{eq:djdrive}.
\begin{table}[b!]
\centering
\setlength{\tabcolsep}{10pt}
\begin{tabular}{ccc}
\toprule
$N$ & \# Osc. Freq ($D_N$) & \# Osc. Freq ($C_s$) \\
\midrule
2 & 0 & 0 \\
3 & 0 & 1 \\
4 & 2 & 4 \\
5 & 4 & 16 \\
\bottomrule
\end{tabular}
\vspace{0.5em}
\caption{Number of oscillation frequencies of long-living correlations for the system of $N$ atoms described by Eq.~\eqref{eq:me} with dipole-dipole interaction Eq.~\eqref{eq:dd_nn}, with respect to the groups $D_N$ and $C_s$.}
\label{tab:osc_pairs_dn_c2}
\end{table}

Since not all of the states from Eq.~\eqref{eq:rho_drive} have a zero real part in the Liouvillian spectrum for $\Delta \neq 0$, they also appear to have a non-zero imaginary part that leads to the oscillations of the corresponding correlation function. The number and degeneracy of oscillating solutions can also be rigorously obtained by the symmetry analysis based on the group representation theory, which is summarized in Table~\ref{tab:osc_pairs_dn_c2}. The state in the form of the Eq.~\eqref{eq:rho_drive} will be oscillating if $\nu$ and $\nu'$ are transformed according to the different representations. Therefore, the number of the oscillating frequencies is given by $\sum_j C_{n_j}^2$ where $n_j$ is the total number of representations for the given $j$ in the related symmetry group (see Table~\ref{tab:repr} in Appendix A). We prove the representation group theory predictions using exact Liouvillian diagonalization in Fig.~\ref{fig:dipole-dipole}(c). This state has a finite, but parametrically small real part compared to $\gamma$; therefore, it gives rise to the non-trivial long-living oscillating dynamics.

{\it Summary.}  In this Letter, we demonstrate the persistence of the long-living collective states in driven atomic ensembles where the collective behavior could be naively expected to be suppressed by the disorder, perturbing the Dicke model. These states are seen as special points of the Liouvillian with a small real part compared to the dissipation.
We show that the driven model can be decoupled from the inhomogeneous broadening for a drive sufficiently larger than dissipation so that these points will enter the kernel of the Liouvillian superoperator.   We also demonstrate that the {dipole-dipole} interaction, if any, can not be completely suppressed by strong driving. Residual dipole-dipole interaction leads to the finite correlation lifetime, and it can also induce non-trivial oscillations in time dynamics. 

In addition to atomic ensembles, the effect could also be applied to ensembles of nuclear spins in solid-state platforms~\cite{KuruczNucSpin}.   In these same platforms, experimental investigations of the time crystal phase are actively ongoing~\cite{greilich_robust_2024}, and, concurrently, the formation of collective dark states has been recently observed~\cite{kirstein_squeezed_2023}.
It could also be instructive to examine the role of disorder and the existence of long-living time-dependent correlations in the waveguide quantum electrodynamics setup~\cite{brehm2020waveguide,poshakinskiy_dimerization_2021,poddubny2023RevModPhys,ShiPoddubny2024}, where the emitters are coupled to a continuum of modes rather than a single mode.

\begin{acknowledgements}
We thank  Boris Altshuler, Nir Davidson, Ziv Meir, Oren Raz, Ephraim Shahmoon, Yoav Shimshi, Dmitry Smirnov and Gianluca Teza for useful discussions. NL acknowledges financial support
from the Israel Science Foundation (ISF), 
The Minerva Foundation, with funding from the Federal German
Ministry for Education and Research.  The work of
ANP has been supported by research grants from the
Center for New Scientists and from the Center for Scientific
Excellence at the Weizmann Institute of Science, and by the
Quantum Science and Technology Program of the Israel
Council for Higher Education. This research is made possible
in part by the historic generosity of the Harold Perlman
Family.

\end{acknowledgements}
\bibliography{PaperLiouvillian_NLvers.bbl}

\clearpage
\let\oldsec\section

\renewcommand{\theequation}{S\arabic{equation}}
\renewcommand{\thefigure}{S\arabic{figure}}
\renewcommand{\thetable}{S{\arabic{table}}}

\setcounter{page}{1}
\setcounter{section}{0}
\setcounter{equation}{0}
\setcounter{figure}{0}
\setcounter{table}{0}

\section*{Supplementary Material}

\subsection*{Appendix A: Symmetry analysis of dark states}\label{app:A}

\begin{figure}[ht!]
    \centering
    \includegraphics[width=0.99\columnwidth]{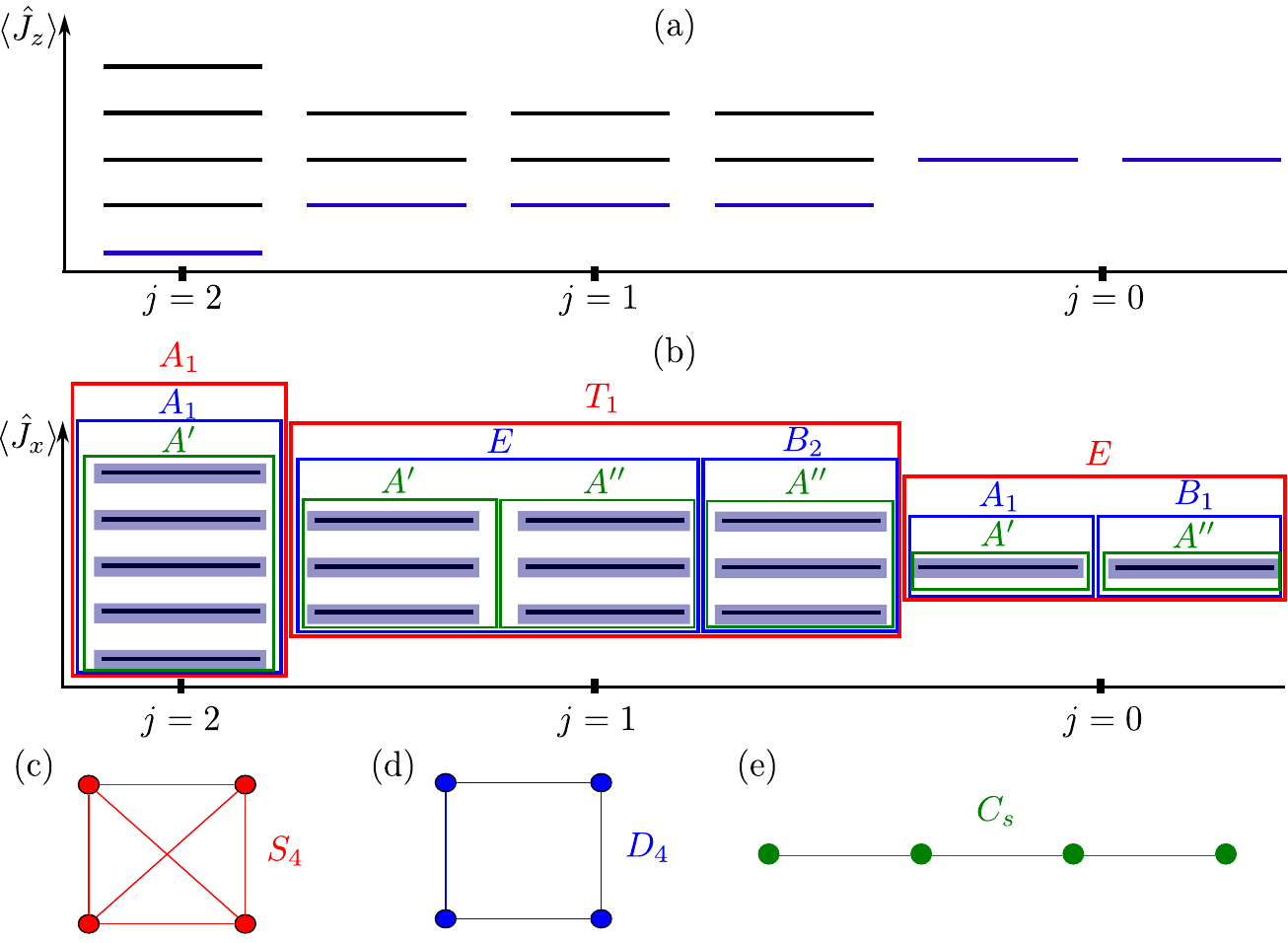}
    \caption{(a) Generalized Dicke ladder for $N = 4$. Each line represents the Dicke state $\ket{j,m,\nu}$ with $j = 2,1,0$; $\langle \hat{J}_z \rangle = m = -j,..,j$ and $\nu = 1,..,d_j$. The dark states in this case are the kernel of the operator $\hat{J}_-$ shown by dark blue lines (see also Fig.~\ref{fig:bloch}(a) of the main text). (b) Generalized Dicke ladder quantized along $\hat{J}_x$ that is the eigenstates of the drive Hamiltonian. Color frames show the irreducible representations of the ladder for 3 symmetry group cases: $S_N$ symmetry illustrated in panel (c) correspond to the system without dipole-dipole interaction when all $N$ atoms are coupled identically (red frames); $D_N$ symmetry illustrated in panel (d) -- system with nearest-neighbor dipole-dipole interaction with periodic boundary conditions (blue frames); $C_s$ symmetry illustrated in panel (e) -- system with nearest-neighbor dipole-dipole interaction with open boundary conditions (green frames). Names of the representations are taken from the Table~\ref{tab:repr}.}
    \label{fig:ladder}
\end{figure}

We start a step-by-step analysis of the dark states with the non-driven case studied in~\cite{Damanet2016,poshakinskiy_dimerization_2021}, see Fig.~\ref{fig:ladder}(a). The number of dark states is then defined solely by the degeneracy of each $j$ manifold~\cite{shammah_superradiance_2017} 
\begin{equation}
    d_j = \frac{N! (2j+1)}{(N/2+j+1)!(N/2-j)!}.
\end{equation}
The total number of dark states is then given by $\sum_j d_j = C_N^{[N/2]}$~\cite{poshakinskiy_dimerization_2021}.

In the case when any coherent processes are present in the system, i.e. ${\cal H} \neq 0$, the long living correlation $\rho_{\rm dark}$ should obey $[{\cal H},\rho_{\rm dark}] = 0$ in order to have ${\cal L}\rho_{\rm dark} \to 0$. It implies that $\rho_{\rm dark}$ could be written in the form 
\begin{equation}\label{eq:alpha}
    \rho_{\rm dark} = \sum_\alpha c_\alpha \ket{\alpha}\bra{\alpha}\:, \quad \text{where} \quad {\cal H}\ket{\alpha} = E_\alpha \ket{\alpha},
\end{equation}
that is diagonal in the basis of eigenstates of $\mathcal H$.
In this reduced basis, the full master equation,  Eq.~\eqref{eq:me} can be rewritten as a rate equation on the coefficients $c_\alpha$ as 
 \begin{equation}\label{eq:kin_eq}
    \dot{c}_\alpha= \gamma\qty(\sum_\beta \abs{\mel{\alpha}{\hat{J}_-}{\beta}}^2c_\beta -c_\alpha \mel{\alpha}{\hat{J}_+\hat{J}_-}{\alpha}).
\end{equation}
 In the case of the strong drive ${\cal H} = 2 \Omega \hat{J}_x$ and $\ket{\alpha} = \ket{j,m_x,\nu}$ are the states quantized along $x$ as shown in Fig.~\ref{fig:ladder}(b). Substituting them into the rate equation~\eqref{eq:kin_eq}, we obtain the answer given in Eq.~\eqref{eq:rho_drive} in the main text. It is important to note that Eq.~\eqref{eq:kin_eq} is relevant only in the case of the single dissipation term $\hat{J}_-$ in Eq.~\eqref{eq:me}. In the more general case of many dissipator operators, it should be modified to the sum of each operator with its own decay rate. 

Generally, the number of linearly independent solutions of the Eq.~\eqref{eq:kin_eq} could be found using the group representation theory. At first, we note, that if the Hamiltonian ${\cal H}$ exhibits any symmetry with respect to the permutation of atoms, e.g., $S_N$ group symmetry (system is permutationally invariant), $D_N$ group (system has a symmetry of $N$ sided polygon), or $C_s$ group (system has a reflection plane), the eigenstates $\ket{\alpha}$ from Eq.~\eqref{eq:alpha} transform according to the representations of this group~\cite{bir1974symmetry}. 

The second insight from the group representation theory that we use is the selection rules, i.e., the vanishing or non-vanishing of the matrix elements due to the symmetry constraints. The theory suggests, that for the matrix element $\mel{\psi}{\hat{V}}{\phi}$ to be non zero it is required that the product of the representation ${\cal D}_{\psi} \otimes {\cal D}_{\phi}$ that correspond to the functions $\ket{\psi}$ and $\ket{\phi}$, respectively, contain the representation ${\cal D}_V$ that correspond to the operator $\hat{V}$~\cite{bir1974symmetry}.

\begin{table}[t!]
\centering
\begin{tabular}{cccccc}
\toprule
$N$ & $j$ & $d_j$ & Rep $S_N$ & Rep $D_N$ & Rep $C_s$ \\
\midrule
\multirow{2}{*}{2} 
  & $1$   & $1$ & $A$ & $A$ & $A'$ \\
  & $0$   & $1$ & $B_1$ & $B_1$ & $A''$ \\
\midrule
\multirow{2}{*}{3} 
  & $\tfrac{3}{2}$ & $1$ & $A_1$ & $A_1$ & $A'$ \\
  & $\tfrac{1}{2}$ & $2$ & $E$ & $E$ & $A'\oplus A''$ \\
\midrule
\multirow{3}{*}{4} 
  & $2$   & $1$ & $A_1$ & $A_1$ & $A'$ \\
  & $1$   & $3$ & $T_1$ & $B_2\oplus E$ & $A'\oplus 2\,A''$ \\
  & $0$   & $2$ & $E$ & $A_1\oplus B_1$ & $A'\oplus A''$ \\
\midrule
\multirow{3}{*}{5} 
  & $\tfrac{5}{2}$ & $1$ & $A$ & $A_1$ & $A'$ \\
  & $\tfrac{3}{2}$ & $4$ & $T$ & $2E$ & $2\,A'\oplus 2\,A''$ \\
  & $\tfrac{1}{2}$ & $5$ & $H$ & $A_1\oplus 2E$ & $3\,A'\oplus 2\,A''$ \\
\bottomrule
\end{tabular}
\caption{Decomposition of the fixed-\(m\) subspace for \(N\) atoms under the permutation group \(S_N\), the dihedral group \(D_N\), and the chain mirror group \(C_s\).. Note that $S_{2,3}$ are isomorphic to $D_{2,3}$, respectively. $S_{4}$ isomorphic to the octahedral group $O$. Names of the representations for the groups $D_N$, $C_s$, and $O$   are given according to~\cite{GelessusCharacterTables2025}. For $S_5$ we use $A$ for the trivial representation, $T$ for the standard representation, and $H$ for the irreducible five-dimensional representation.}
\label{tab:repr}
\end{table}

\begin{figure*}[ht!]
    \centering    \includegraphics[width=0.99\linewidth]{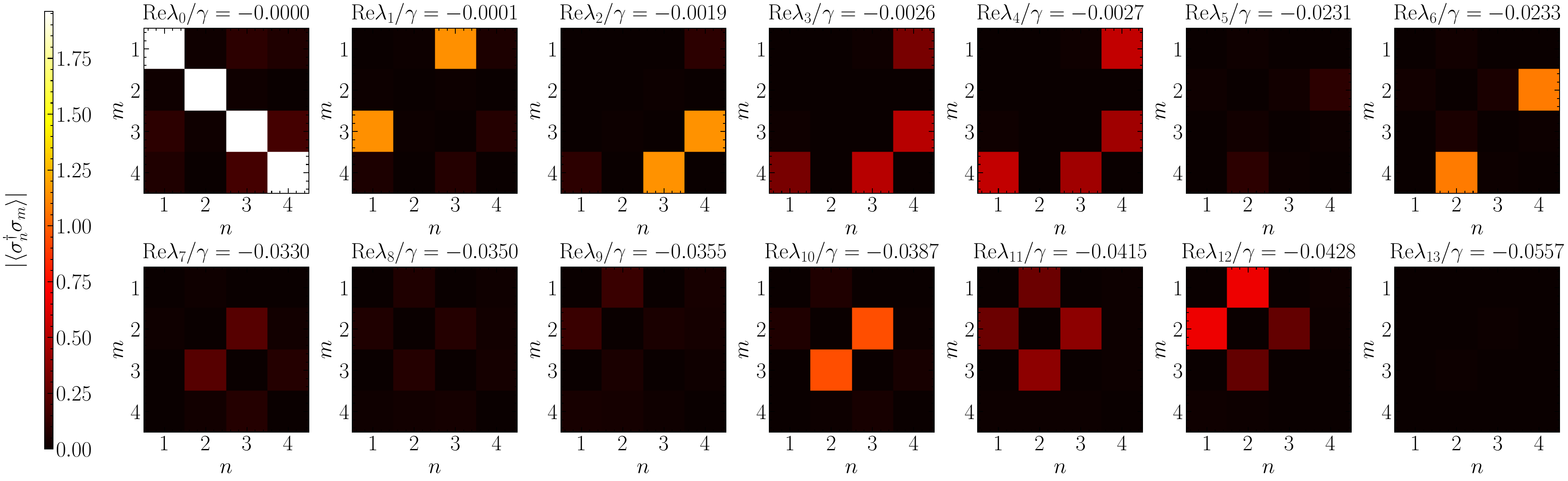}
    \caption{Contribution to the correlation functions $|\langle \sigma_n^\dagger 
    \sigma_m \rangle|$ from the eigenvectors corresponding to the first smallest absolute value eigenvalues of the Liouvillian given by Eq.~\eqref{eq:me} in the main text calculated for the same parameters as in Table~\ref{tab:exp_decay}.}
    \label{fig:eigenvecs}
\end{figure*}

In this Letter, we analyze three cases of the Hamiltonian symmetry, shown in Fig.~\ref{fig:ladder}(c-e) for $N = 4$ atoms. The case of the $S_N$ symmetry corresponds to the Hamiltonian ${\cal H} = 2\Omega\hat{J}_x$ and is realized in the model Eq.~\eqref{eq:me} at large drive $\Omega \gg \gamma,\delta \omega$, since, as we have shown in the main text, the large drive decouples the system from the disorder, that is the $\delta H$ term in Eq.~\eqref{eq:me}. The Hamiltonian that also includes the dipole-dipole nearest neighbors coupling given by Eq.~\eqref{eq:dd_nn} obeys $D_4$ or $C_s$ symmetry depending on the atom's location as shown in Fig.~\ref{fig:ladder}(d-e).

We have analyzed the representations that appear in the Dicke ladder for all these groups in Table~\ref{tab:repr}. Regarding the Eq.~\eqref{eq:kin_eq} the operators $\hat{J}_-$ and $\hat{J}_+\hat{J}_-$ are symmetric sums of the single atoms operators, so they transform according to the trivial representation of each group, i.e. $A$ in $S_N$ and $D_N$, and $A'$ in $C_s$. With that, we conclude that the non-trivial solutions (when matrix elements are not zero) of Eq.~\eqref{eq:kin_eq} appear only if $\ket{\alpha}$ and $\ket{\beta}$ are from the same representation. Then, the total number of solutions of the equation $\dot{c}_\alpha = 0$, which  also includes the steady state, is given by 
\begin{equation}\label{eq:sumDN}
    \Sigma = \sum_{{\cal D}_i} |\text{dim}~{\cal D}_i|^2,
\end{equation}
where ${\cal D}_i$ are the irreducible representations that describe the states in Fig.~\ref{fig:ladder}(a-b).

\subsection{Appendix B: Relation between Liouvillian eigenvectors and correlations $\langle \sigma_n^\dagger \sigma_m \rangle $}\label{app:B}

In order to prove that long-living correlations are directly related to the eigensystem of the Liouvillian at the strong drive, we calculate the observable $\langle \sigma_n^\dagger \sigma_m \rangle $ for eigenvectors that correspond to the 14 smallest eigenvalues of the Liouvillian in Fig.~\ref{fig:eigenvecs}. According to the Eq.~\eqref{eq:L_eig} the real part of each eigenvalue defines the decay rate of the correlation function its eigenvector contributes to. Then one could compare it with the direct exponential fit of the time dynamics simulation with the same parameters given in Table~\ref{tab:exp_decay}.  

\begin{table}[b!]
\centering
\begin{tabularx}{\columnwidth}{>{\centering\arraybackslash}X>{\centering\arraybackslash}X>{\centering\arraybackslash}X>{\centering\arraybackslash}X}
\multicolumn{2}{c}{$|\langle \sigma_n^\dagger \sigma_m \rangle|$} & \multicolumn{2}{c}{Exponential fit $Ae^{-Bt}$} \\
\toprule
$n$ & $m$ & $A$ & $B/\gamma$ \\
\midrule
1 & 2 & 0.1813 & 0.0459 \\
1 & 3 & 0.1667 & 0.0002 \\
1 & 4 & 0.1676 & 0.0035 \\
2 & 3 & 0.1800 & 0.0422 \\
2 & 4 & 0.1737 & 0.0231 \\
3 & 4 & 0.1675 & 0.0030 \\
\bottomrule
\end{tabularx}
\caption{Exponential fits for the correlation functions $|\langle \sigma_n^\dagger \sigma_m \rangle|$ calculated at $\Omega = 200\gamma$ for the random frequency disorder generated from the normal distribution with the scale $10\gamma$: $\omega_1/\gamma = -0.62448819$, $\omega_2/\gamma = 5.93539815$, $\omega_3/\gamma = -1.53186917$, $\omega_4/\gamma = 3.04670911$ (same parameters as in Fig.~\ref{fig:time_dyn} in the main text).}
\label{tab:exp_decay}
\end{table}

Comparing Figure~\ref{fig:eigenvecs} and Table~\ref{tab:exp_decay}, we see a clear relation. Indeed, for example, while the eigenvector that 
corresponds to the eigenvalue $\lambda_1$, which is the smallest nonzero eigenvalue has a significant contribution to the correlation $|\langle \sigma_1^\dagger \sigma_3 \rangle|$ as seen in Fig.~\ref{fig:eigenvecs}, the time dynamics of the correlation simulations show that $|\langle \sigma_1^\dagger \sigma_3 \rangle|$  has the biggest lifetime same order as $\lambda_1$. The same correspondence is clearly seen between the correlation function $|\langle \sigma_2^\dagger \sigma_4\rangle|$ and $\lambda_6$. For the rest of the correlation functions listed in Table~\ref{tab:exp_decay}, we see from Fig.~\ref{fig:eigenvecs} that they have contributions from different eigenvectors, so we can not claim the exact eigenvalue for them. However, their decay rate has the same order as the smallest eigenvalue, which the eigenvector contributes to it: $|\langle \sigma_3^\dagger \sigma_4\rangle|$ as $\lambda_2$, $|\langle \sigma_1^\dagger \sigma_4\rangle|$ as $\lambda_3$, $|\langle \sigma_2^\dagger \sigma_3\rangle|$ as $\lambda_7$, $|\langle \sigma_1^\dagger \sigma_2\rangle|$ as $\lambda_{11}$.     

\end{document}